\documentstyle[12pt]{article}
\begin{document}
\thispagestyle{empty}
 \noindent
\hspace*{115mm} hep-th/9512060 \\
\hspace*{115mm} DTP-95-77\\
\hspace*{115mm} TUW-95-25\\
\hspace*{115mm} November 1995\\
\begin{center} \vspace*{5mm}
\Large \bf More Dual String Pairs From Orbifolding\\

\vspace*{15mm} \large H. B. Gao$^\star$ \\
\normalsize {\it Department of Mathematical Sciences, University of Durham\\
South Road, Durham DH1 3LE, UK}\\
and\\
\normalsize{\it Institut f\"ur Theoretische Physik, TU Wien$^\dagger$\\
Wiedner Hauptstr. 8-10, A-1040 Wien, Austria}\\

\vspace*{35mm} \large Abstract \\ \end{center}

We construct more dual pairs of type II-heterotic strings in four
dimensions with $N=2,1$ spacetime supersymmetry. On the type II
side the construction utilizes the various possible choices of K3
automorphisms with fixed points which transform the holomorphic
two-form nontrivially, and rotation plus translation
on $T^2$. The Calabi-Yau orbifolds so obtained have non-zero Euler
numbers, so quantum corrections exist on the type IIA strings. The
heterotic string (asymmetric) orbifold duals are found which depend on
going to the enhanced symmetry points. Some aspects of the construction
are discussed including the role of the singularity and the possibility
of going beyond the adiabatic argument. Many of these examples have
also orientifold analogs.

\begin{flushleft}
\vspace*{15mm}
$\star$ {\small On leave from Zhejiang
Institute of Modern Physics, Zhejiang University, Hangzhou, China}\\
$\dagger$ {\small Present address.} \end{flushleft}

\newcommand{\be}{\begin{equation}}
\newcommand{\ee}{\end{equation}}

\newpage
\section{\bf Instroduction}

Since the work of Sen and Harvey and Strominger \cite{sen, hs} on
the conjectured
equivalence of the type IIA string compactified on K3 and the heterotic
string compactified on $T^4$ \cite{hult},
string-string duality  has been
a topic of a large amount of recent literature \cite{kacvaf, 2ndqu,
witva, senva, n1, aspmor, aspin2, aspin1, chaulow, afiq}.
Among the consequences of this conjecture, four dimensional dual string
pairs constructed from compactifying further on $T^2$ and orbifolding
 \cite{witva, 2ndqu} are of  special
interests. On the one hand, it is noted in \cite{2ndqu} that due to
low energy decoupling of $N=2$ vector and hypermultiplets, models can be
constructed which, by using duality, are capable of producing exact
calculations on the quantum moduli space of vector and hypermultiplets
on both side. On the other hand, there is indication \cite{kac2}
that the point particle
limit of some of the four dimensional dual string pairs constructed in
\cite{kacvaf} may reproduce the main points of the exact results in
the Seiberg-Witten theory \cite{seiwit}. Therefore, it seems likely
that a  complete
understanding of the four dimensional string duality is an important
step towards formulating  new string theory and deducing its
phenomenological consequences.

Despite of the impressive progress, there remain still quite a lot of
problems to be solved. For example, we do not know how to deduce the
results
of \cite{kacvaf} from the string-string duality in six dimensions
\cite{sen, hs} which is better understood than any other analogous
conjectures. Initial attempt has been taken in \cite{witva} by using
adiabatic argument, and by exploring interesting but still mysterious
structure of K3-fibration \cite{klm,klt, liany}
arising in all of the Calabi-Yau hypersurfaces
used to compactify type II string duals of \cite{kacvaf}.
In \cite{afiq}, certain inter-relationship between Kachru-Vafa models
is examined from the heterotic side. In \cite{ asplo}, it is confirmed
that K3-fibration structure is consistent with four dimensional
type IIA-heterotic string duality. Other problems include the origin of
the enhanced gauge symmetry in the type II dual formulation, which has
been actively explored by Aspinwall \cite{aspin3, aspin4} recently.

In trying to generalize the orbifold construction of \cite{2ndqu} to
get more dual pairs, instead of using fixed point free K3 automorphisms
which are rather rare, one uses automorphisms which act with fixed points
on the K3 factor but act freely on the product K3$\times T^2$ so that the
quotient is smooth at least away from the boundary components of the moduli
space. In this way, one usually obtains asymmetric orbifolds on the
heterotic side, often at the specific point in moduli space with enhanced
gauge symmetry. According to \cite{aspin4,aspin1}, the type IIA
dual of the enhanced symmetry point corresponds to the degenerate
K3 or Calabi-Yau spaces, non-perturbative phenomena such as transition
between different phases (topologies) is visible by studying the details
of these degenerate points. Thus dual string pairs at the enhanced
symmetry point are useful also in understanding phase transition.
Perhaps most importantly, non-trivial instanton corrections are present
on the heterotic side whose exact form may be deduced from the
world sheet instanton corrections on the type IIA side by duality.

In \cite{chaulow, swsen, aspin2, n1} asymmetric orbifolds are
constructed in 4 and higher dimensions by
 utilizing the  list of the K3 automorphisms
\cite{niku} with fixed points. As the above references deal with
dual pairs with maximal spacetime supersymmetries, the K3 automorphisms
keeping holomorphic two-form invariant are considered. From many
phenomenological points of view, string duality in realistic dimensions
and with $N=2$ or less supersymmetry may be more interesting. One way to
achieve this, as we will do in this letter, is by using K3 automorphisms
which change the holomorphic two-form on K3, but leave invariant
a holomorphic three-form on the product K3$\times T^2$, thus the quotients
are smooth Calabi-Yau manifolds on which type IIA string is compactified.
The Euler numbers of the Calabi-Yau spaces are different from zero, thus
it is probable that there exist worldsheet instanton corrections to the
type II moduli space which are equivalent to the spacetime instanton
effects under duality. In this letter we will only construct various
possible dual pairs and will leave the problems of calculating instanton
corrections as well as analyzing phase transitions untouched. It is
certainly worthwhile to pursue these questions along the lines of \cite{
2ndqu, aspin1}.

\section{\bf Construction of Dual Pairs}

The basic procedure of orbifolding to get dual pairs in
4 dimensions has been outlined in \cite{2ndqu, witva, n1}.
The strategy is to start from six dimensional type II-Heterotic
string duality, and further compactify both side on $T^2$. Upon
performing various quotient by discrete groups of the K3
automorphism on the type II side, together with rotations and
translations of the $T^2$, one obtains smooth Calabi-Yau spaces on
the type IIA side. Heterotic string duals are obtained by examining
how the discrete groups act on the cohomology lattice $H^2(K3\times
T^2, Z)$ which is identified \cite{hs} as the heterotic Narain
lattice $\Gamma^{22,6}$. For discrete K3 automorphisms of order
higher than two, this is achievable by grouping the root lattices
of certain Lie algebras  embedded in $\Gamma^{19,3}$
according to the fixed point sets of the respective cyclic groups.
For the case of $Z_2$ however, there are fewer choices. We happen
to find a $Z_2$ automorphism which acts on the K3 surface in a way
that its lift to the cohomology lattice is obtainable by embedding
of the Enriques lattice with finite automorphism group. The heterotic
duals then follow from standard identification of moduli spaces of
vector and hypermultiplets of the $N=2$ heterotic string with the invariant
and anti-invariant sublattices.
As usual, level-matching sometimes requires shift of certain lattice
vectors of the heterotic asymmetric orbifolds.

\vskip 4mm
\noindent{\large \it $Z_2$ Revisited}
%\vskip mm

Our first example comes from modding the $N=4$ pair
of type IIA on K3$\times T^2$, and heterotic string on $T^4\times
T^2$, by $Z_2$.

On the type II side, consider the K3 represented as the qaurtic
hypersurface in $CP^3$
\be z^4_1 +z^4_2+z^4_3+z^4_4=0, ~~ z_i \in CP^3, \label{k3:eq} \ee
the holomorphic (2,0) form is given by
\be
\omega_{(2,0)} =\Bigl(\frac{\partial P}{\partial
(z_4/z_1)}\Bigr)^{-1} d\Bigl(\frac{z_2}{z_1}\Bigr)\wedge
d\Bigl(\frac{z_3}{z_1} \Bigr)
=\frac{1}{4z^3_4}(z_1dz_2\wedge dz_3+z_3dz_1\wedge dz_2+z_2dz_3\wedge
dz_1). \label{k3:2fm} \ee
$P$ in (\ref{k3:2fm}) is the defining polynomial of (\ref{k3:eq}) in
homogenious variables.
It is easily seen that the following K3  automorphism\footnote{It may
seem unusual that this automorphism contains a permutation of coordinates.
One may however convince oneself that this kind of the K3 symmetry
in fact exists in at least a 4 dimensional subspace of the 20 dimensional
K3 moduli space, i.e., those arising from polynomial perturbations
$\alpha z_1 z_2 z_3 z_4, ~\beta z^2_2 z^2_3, ~\gamma z_2 z_3 z^2_4, ~
\delta z^2_1 z_2 z_3$. Given that at the moment the string-string duality
admits still "point spectrum", there is no  doubt that this symmetry,
though existed in a moduli subspace, is worth considering.}
\be g:~(z_1, z_2, z_3, z_4)\longmapsto (-z_1, z_3, z_2, -z_4),
  \label{k3:aut1}\ee
transforms $\omega_{(2,0)}$
\be g^*\omega_{(2,0)}=-\omega_{(2,0)}. \ee
And for the $Z_2$ action on $T^2=S^1_a \times S^1_b$, we choose the
reflection in both factors and a translation on one of the circles,
e.g. $S^1_a$:
\be g: (x,y)\longmapsto (-x+\frac{1}{2}, -y). \ee
or in the complex coordinate $w=x+iy$  of $T^2$:
\be g: w \longmapsto -w+ \frac{1}{2}. \label{t2}\ee

Note that the $Z_2$ action in (\ref{k3:aut1}) has four fixed points on K3
but the action on the product K3$\times T^2$ is fixed point free.
Without the translation in one of the circles, the total reflection
would have four fixed points on the torus $T^2$. Since the action on
K3 is not free, we have to make the translation in order for the
quotient to be smooth. It is worthnoting that in the large  radius limit
$R_a \rightarrow \infty$, one recovers two fixed points on $T^2$, and
more complicated fixed sets on K3$\times T^2$.

A variant of the above model for K3$\times T^2/Z_2$ can be constructed by
considering orientation-reversing involutions on $T^2$. There are two
possibilities, either by modding out symmetry
\be w \longmapsto \bar{w}+\frac{1}{2} \label{ori1} \ee
or \be w \longmapsto -\bar{w} +\frac{i}{2} \label{ori2}\ee
both leading to smooth quotients. In the large radius limit for
either of circle $a$ or $b$, one recovers a singular curve, instead of
singular point.

To count the states of this orbifold,
we must specify the action of this K3 automorphism on the space of
cohomology of K3$\times T^2$.
{}From the Lefshetz fixed point formula, it follows that
\be 4=2-2+tr~g|_{H^{1,1}}, \ee
where 4 is the number of fixed points, the first 2 on the right comes
from invariant (0,0)- and (2,2)-forms, the $-2$ from the (2,0)-and
(0,2)-forms which are odd under (\ref{k3:aut1}). Thus we have the
eigenvalues of the automorphism $g$ on the 20 (1,1)-forms on K3, $((-1)^8,
(1)^{12})$. Combined with the invariant (1,1)-form on $T^2$, this gives
$h^{1,1}=13$. For the (2,1)-forms, we have one from combining a holomorphic
(2,0) form on K3, $\omega_{(2,0)}$ with a (0,1)-form from the torus which
are both $Z_2$-odd, and 8 from the odd (1,1)-forms on K3. Therefore
$h^{21}=9$. Notice that the resulting Calabi-Yau orbifold has nonvanishing
Euler number, in contrast to the case of \cite{2ndqu}.

In order to find the  heterotic dual of this model, we have to first
understand  how the $Z_2$-automorphism acts on the cohomology lattice
$H^2(K3\times T^2, Z)$. Then
using the identification of K3 cohomology lattice with
the Narain lattice of the toroidal compactification of heterotic string
 \cite{sen, hs, aspmor}, the $Z_2$ action of (\ref{k3:aut1}) is
translated into heterotic aymmetric orbifold.
Note that the K3 automorphism (\ref{k3:aut1}) looks similar to half of
an automorphism used to obtain Enriques surface, combined with the one
with eight fixed points. Actually K3 can be viewed as the covering space
of the Enriques surface.
It is thus natural to use finite automorphisms
of the Enriques lattice embedded in the K3 lattice to represent the
action of (\ref{k3:aut1}) on the  K3 cohomology lattice.
The necessary techniques are explained in
\cite{bart, niku}. In fact, there exist on Enriques lattice containing
rational curves certain root invariants formed by the lattice
vectors obeying $q^2=-2$.
Using two of the six root invariants of the Enriques
moduli space listed in \cite{niku}, namely $(D_5\oplus D_5,\{0\})$
and $(E_7 \oplus A_3, Z_2 )$ where the groups $\{0\}, Z_2$ denote
the kernels of the homomorphism
between Picard groups of K3 and Enriques surface, and the embedding
of the Enriques lattice into the K3 lattice:
\be E_8 \oplus H \rightarrow E_8 \oplus E_8 \oplus H\oplus H \oplus H
\label{emb}\ee
one can determine the $Z_2$ action on $H^2(K3, Z)$, via the above
embedding, as interchanging the two $D_5$ factors and reversing the sign
of the $A_3$ factor. This can be summarized as follows.
One can decompose the
cohomology lattice $H^2(K3\times T^2, Z)$ into even lattices
\be \Gamma^{22,6}=\Gamma^{7,1}\oplus\Gamma^{3,1}\oplus\Gamma^{5,1}_a
\oplus\Gamma^{5,1}_b\oplus\Gamma^{1,1}_1\oplus\Gamma^{1,1}_2,
\label{z2:lat1} \ee
where the last two terms correspond to the  $T^2$.
Note that the K3 parts in the decomposition (\ref{z2:lat1}) are not
self-dual, this is related to the fact  that Enriques lattice
can not be embedded into K3 lattice as a direct factor.
Let $\gamma^*$ be the corresponding vector in the lattice $\Gamma^*$,
then the action is
\be g|\gamma^{7,1}, \gamma^{3,1},\gamma^{5,1}_a,\gamma^{5,1}_b,
\gamma^{1,1}_1, \gamma^{1,1}_2\rangle
=|\gamma^{7,1},-\gamma^{3,1}, \gamma^{5,1}_b, \gamma^{5,1}_a,
-\gamma^{1,1}_1, -\gamma^{1,1}_2\rangle, \label{z2:lat}\ee
where we assign a twist by null vector in the last direction.
It is perhaps worth emphasizing that the above lattice decomposition
(\ref{z2:lat1}) is not self-dual, thus corresponds not to the
generic point in Narain moduli space. This is so, what we obtain
is a dual heterotic orbifold at enhanced symmetry point.
The action $g$ has ten -1 eigenvalues on the left and four -1
eigenvalues on the right. In particular, the right moving
heterotic coordinates corresponding to the holomorphic (2,0)-form
of the type II side are projected out, thus the supersymmetry is
reduced  in this model.  It can be shown that the shift of zero
point energy on the left is nonzero, thus level-matching requires
an additional shift vector be added. This vector $\delta=(p_L,p_R)/2$
must satisfy $p^2_L -p^2_R=3$.
The massless spectrum consists of ten hypermultiplets and thirteen
 vector multiplets, in addition to the gravitational one.

At this point it is interesting to compare with the result of \cite{
2ndqu}. We obtain the Calabi-Yau on the type IIA side which is not
self-mirror, i.e. $\chi \not= 0$. This must mean that there exist
non-perturbative instanton corrections on the type II side which
maps to the spacetime instanton corrections on the  heterotic
side. Further analysis is needed in order to use the (first quantized)
mirror map to calculate the instanton correction on the heterotic side,
this could provide a
nontrivial test of the 2nd quantized mirror symmetry proposed in \cite{
2ndqu}.

\vskip 4mm
\noindent {\large \it Other Cylic Groups}
%\vskip 3mm

The analysis for the other cyclic groups can be made
in the same fashion. For $G=Z_p, ~~p=3,5,7$, the action of $G$
on the K3 is by multiplying group elements $\exp (2ki\pi /p),
0\le k< p$ to the appropriate complex variables $z_i,~~ i=1,..,4$
of some weighted $CP^3$ in which the K3 is realized.
And on the torus $T^2$, the $G$ action is such that rotation on
one of the circles accompanies the translation in order to compensate
for the nontrivial transformation of the holomorphic form on K3
while retaining fixed point free action on the product K3$\times
T^2$. For example, take $G=Z_3$, one notices that the K3
automorphisms
\be g_1: ~~(z_1, z_2, z_3, z_4)\mapsto (e^{2i\pi/3}z_1, e^{2i\pi/3
}z_2, z_3, z_4),\ee \be
g_2: ~~(z_1, z_2, z_3, z_4)\mapsto(z_1, z_2, e^{2i\pi/3}z_3,
e^{2i\pi/3}z_4)\ee
of $Z_3$ transform the holomorphic two-form as
\be g^*_1\omega_{(2,0)}=e^{2i\pi/3}\omega_{(2,0)},\ee\be
g^*_2\omega_{(2,0)}=e^{-2i\pi/3}\omega_{(2,0)}\ee
which can be rendered invariant if one tensors it
with $dw$ from the
$T^2$ with nontrivial $Z_3$ transformations.
The number of fixed points under either $g_1$ or $g_2$ is nine;
it is six under both. Counting of states can be carried out similarly.
For example, from the untwisted sector we have two invariant
 (1,1)-forms, one of them is the K\"ahler form.
There are eighteen anti self-dual (1,1)-froms which are cyclically
interchanged by $g_1, g_2$, thus by forming linear combinations six
out of them are also $Z_3$-invariant. Unlike the
$Z_2$ case, now a $Z_3$ transformation does not leave the (1,1)-form
from $T^2$ invariant (the same is true for other higher order cyclic groups).
So we obtain $h^{11}=8$. The invariant (2,1)-forms can be counted
as follow. There is one coming from tensoring $\omega_{(2,0)}\otimes
d{\bar w}$ on K3$\times T^2$. From the twisted sector, we get six
(1,1)-froms which when tensored by $dw$ from $T^2$ are $Z_3$-invariant.
Since there are two twisted sectors, the final result is $h^{21}= 13$.

The heterotic dual of this orbifold can be realized as an
asymmetric orbifold with enhanced gauge symmetry similar to that of
\cite{chaulow, aspin2}.
 We consider the decomposition of the Narain lattice as follows:
\be \Gamma^{22,6}=E_8\oplus E_8\oplus \Gamma^{2,2}_1\oplus
\Gamma^{2,2}_2\oplus \Gamma^{2,2}_3 \label{z3:lat}\ee
which contains the enhanced symmetry of type $(D^6_1)^3\times D^2_1
\times D^4_1$ embedded in the 24 dimensional Niemeier lattice $D^4_6$.
It is then not difficult to specify the $Z_3$-action on (\ref{z3:lat}):
The three $D^6_1$ factors are cyclically interchanged
under $Z_3$. Since the right-moving coordinates
corresponding to the self-dual two-forms on the type IIA side transform
nontrivially, supersymmetry is reduced. Level-matching requires shifting
the lattice $\Gamma^{2,2}_2$ by a vector $\delta^2=\frac{1}{3}$,
in addition to the the null vector shift in the third factor
$\Gamma^{2,2}_3$.

For groups $Z_5, Z_7$, one can go on to find the
Calabi-Yau orbifolds, but now since there would be no room for putting
shift in the lattice, level-matching rules out the existence of
corresponding heterotic duals. Coincidently, the Calabi-Yau manifolds
obtained in these examples have the same Hodge numbers
as those obtained from the examples of $Z_6$ and $Z_8$, respectively.
Thus there is no loss as far as getting heterotic-type IIA dual pairs
is concerned.

For $G=Z_m, ~~m=4, 6, 8$, to construct Calabi-Yau orbifolds, care must
be taken since the actions by the subgroups $Z_2, Z_3$ may be different
from that of the higher order elements. For example, $Z_4$ has two
elements of order four, one element of order two, acting by
rotation by $\pi/2$ and reflection, respectively,
on the complex coordinates of the K3.
For the K3 realized as in (\ref{k3:eq}), one would not obtain Calabi-Yau
orbifold by pair-wise action on the coordinates since the order two
element always acts preserving holomorphic two-form $\omega_{(2,0)}$.
Likewise, the choice of action of the $Z_2$-subgroup as in (\ref{k3:aut1})
would not be appropriate since then one is actually modding by the
semi-direct product group $Z_4\times S_4$
which could not be translated
into conventional dual heterotic orbifold. Nevertheless, if one
chooses the $Z_4$-action such that the order four element acts by
multiplying $\omega_{(2,0)}$ by $\pm i$, the order-two element maps it
nontrivially as well. Note that for this $Z_4$ example, there are eight
fixed points for each of the order-four elements, and 12 for the order-two
element. Only 4 of them are common fixed points. Standard counting gives
the Hodge numbers of this $Z_4$ orbifold as $h^{1,1}=6, ~h^{2,1}=15$.
And similarly we have $h^{1,1}=4, ~h^{2,1}=17$ for $Z_6$, and
$h^{1,1}=2, ~h^{2,1}=19$ for $Z_8$. All of these type II Calabi-Yau
orbifolds have (asymmetric) heterotic orbifold duals near
 the enhanced symmetry point. We summarize the result in Table \ref{tab:1}.

\begin{table}[b]
\centerline{\begin{tabular}{|c||c|c|c|c|}
\hline $G$&$\#Fix$(K3)&$n_v=h^{1,1}$&$n_h=h^{2,1}+1$& shift $\delta^2$\\
\hline \hline $Z_2$&4&13&10&3/4\\
\hline $Z_3$&6&8&14&1/3\\
\hline $Z_4$&4&6&16&1/4\\
\hline $Z_6$&2&4&18&5/4\\
\hline $Z_8$&2&2&20&1/8\\
\hline
\end{tabular}}
\caption{Type II Calabi-Yau orbifolds whose heterotic duals can be found.
The second column indicates the number of fixed points on the K3.
The last column shows the shift vectors required to satisfy the
level-matching condition. } \label{tab:1}
\end{table}

\vskip 4mm
\noindent {\large \it Product Groups}
%\vskip 3mm

Now we consider the case of K3$\times T^2/G$, where $G=Z_2\times Z_2,
{}~~Z_2\times Z_4, ~~Z_3\times Z_3, ~~Z_4\times Z_4, ~Z_2\times Z_6$.
These groups have a number of subgroups which fall into the classes
in the previous examples. In general, one obtains $N=1$ spacetime
supersymmetry by suitably choosing the $G$-action on the right-movers.
It is also possible to obtain $N=2$ dual pairs using the
methods of \cite{chaulow}. Here we only consider orbifold with
$N=1$ spacetime supersymmetry, thus we require that holomorphic two-form
on K3 is transformed by elements of both factors.

Let us consider the
modding by group $Z_2\times Z_2$ which  contains
three $Z_2$ subgroups. Denoting their elements as $g_1, g_2$
and $g_3$, respectively,
we take the action of $g_1$ as in (\ref{k3:aut1}) and that of $g_2$
as the "conjugate" of $g_1$ in such a way that the third $g_3=g_2g_1$
acts by even permutations of two pairs of the coordinates, $z_i,~i=1,
... 4$, and without reflections.
There are two subtle points worth noting. Firstly, one cannot
at the same time choose all three of $g_i$ acting on K3 as in
(\ref{k3:aut1}) for reason that $S_4$ has no order two subgroup needed
to make the actions of $g_1$ and $g_2$ commute. Secondly, the action of
the product $g_3=g_2g_1$ necessarily preserves the holomorphic
two-form on K3, as is easily checked. This is consistent with the
transformation of $Z_2\times Z_2$ on $T^2$, where the product of the
two commuting transformations of the type (\ref{t2}) preserves the
one-form $dw$. For the above choice of the $Z_2\times Z_2$ action on
K3$\times T^2$, it is not difficult to see that there are four fixed
points for each of $g_1, ~g_2$, and eight for $g_3$. Examining this
$Z_2\times Z_2$ action on $H^2(K3)$, one obtains the following
spectrum of the type IIA Calabi-Yau orbifold.
There are five (1,1)-forms
surviving the $Z_2\times Z_2$ projection. The other sixteen
anti-self-dual (1,1)-forms are generically interchanged.
Especially spacetime supersymmetry is broken to $N=1$.
Since the K\"ahler mode is left invariant in this case, of the
five $N=2$ vector multiplets which are even under $Z_2\times Z_2$
each contributes an additional chiral multiplet after truncation
to $N=1$. Combined with sixteen truncated hypermultiplets and
three corresponding to the $T^2$ moduli $S, T, U$, this yields
24 chiral multiplets.
We have therefore obtained the field content of a $N=1$ supergravity
coupled to five  $N=1$
vector supermultiplets and 24 massless chiral multiplets.

Now we map this to the $N=1$ heterotic dual. We know the $Z_2\times
Z_2$ action on the cohomology lattice of K3$\times T^2$, which is
then translated into corresponding action on the Narain lattice from the
heterotic side. The eigenvalues of three $Z_2$ elements are shown
below (left and right components are separated by ';' and repeated
eigenvalues are denoted by superscripts)
\be g_1: ~(1^7, (-1)^3, 1^5, (-1)^5, (-1)^2; 1, -1, 1, -1, (-1)^2)\ee
\be g_2: ~(1^5, (-1)^5, 1^7, (-1)^3, (-1)^2; 1, -1, 1, -1, (-1)^2)\ee
\be g_3: ~(1^2, (-1)^8, 1^2, (-1)^8, 1^2; 1,-1,1,-1,1^2).\ee
Note that the product action $g_3$ satisfies level-matching without
additional shifts. The massless spectrum includes five vector multiplets
coming from the untwisted sector. Generically there are no massless
states in the twisted sectors. The invariant projection of the
states $\alpha^I_{-1}|0\rangle_L\otimes |i\rangle_R, ~I=1,...,22,
i=1,...,6$ gives the 24 scalar multiplets including gravitational
state. Note that dilaton is now in a chiral multiplet.
At the enhanced symmetry point the low energy
gauge group is of the form $SO(4)\times SO(4)\times U(1)$ with the last
factor coming from the invariant $N=2$ vector multiplet containing dilaton.
At the self-dual radius of one of the circle in $T^2$, one gets
$SO(4)\times SO(4)\times SU(2)$. The ability of having an $SU(2)$ point
independent of the enhanced symmetry point in the bulk of moduli space
 may be phenomenologically interesting as one can study
low energy effective theory with different field content and gauge
groups by Higgsing along different directions.
Examining the worldsheet currents at the enhanced symmetry point tells
us that the gauge group is at Kac-Moody level four.

Other $N=1$ dual pairs from orbifolding by the product groups $G$ are
similarly worked out with the results summarized in Table \ref{tab:2}.

\begin{table}[b]
\centerline{\begin{tabular}{|c||c|c|c|c|}
\hline $G$&vector&chiral& enhanced points&K-M level\\ \hline
\hline $Z_2\times Z_2$&5&24&$SO(4)\times SO(4)\times SU(2)$&4\\
\hline $Z_2\times Z_4$&3&21&$SU(2)\times SO(4)$&8\\
\hline $Z_2\times Z_6$&2&16&$SU(2)^2$&12\\
\hline $Z_3\times Z_3$&4&19&$SO(8)$&9\\
\hline $Z_4\times Z_4$&2&17&$SO(4)$&16\\
\hline
\end{tabular}}
\caption{$N=1$ heterotic duals to the type IIA on K3$\times T^2/G$.
Indicated here are also possible low energy gauge groups which
appear at the enhanced symmetry points, and their Kac-Moody levels.
 } \label{tab:2}
\end{table}

\section{\bf Comments and Discussions}

All the dual pairs in the preceding section are obtained via standard
orbifold techniques, they appear to parallel the existing dictionary
of the type II-heterotic string duality in four dimensions. Some
special features of this construction are worth emphasizing.

Firstly,
the singularities at the large radius limit in our examples seemed
to invalidate the adiabatic argument of \cite{witva}.
Indeed, fiber-wise application of the six dimensional string duality
requires smooth action of the orbifolding group on the base, for
sufficiently large radius. The presence of quotient singularities
at large radius limit certainly says that there are subtleties in
applying string duality near the boundary of the moduli space.
In four dimensional heterotic string, dilaton comes from the modulus
(area) of the two-torus, thus large radius limit corresponds to weak
coupling of the heterotic string. Appearance of enhanced symmetry in
the weakly coupled limit has been recently observed \cite{aspin4}
at least for the perturbatively visible parts of the group \cite{
asplo}. The question then is, can we prove that taking the weak-coupling
limit commutes with going to the boundary corresponding to the enhanced
gauge symmetry. An affirmative answer clearly saves the adiabatic
arguments from suffering from singularity, at least those appearing
in the large radius limit. Recently, enhanced gauge symmetry points
are examined from the D-string point of view \cite{dstr}, among other
advantages, this enables more geometric description of the degeneration
within Calabi-Yau threefolds. One of the new results is that, instead of
shrinking down to zero size the rational curves, the proccess is better
understood as vanishing of 3-cycles \cite{wit2},
 $S^2_{ij}\times (A,B)$ where $S^2_{ij}$ comes from the singular locus of
 K3, and the $A, B$ are either of the circles of $T^2$. The cycles
containing $A,B$ are never mixed.
It is clear from this picture that one can make the the two proccesses
commute, thus avoiding contradiction with adiabatic argument.

Another point of interests in our construction is the use of the large
radius limit singularity in realizing the extremal transition \cite{aspin1}.
The phase transition usually occurs when
one blows down certain rational curves followed by blowing-ups.
The ability to blow down curves nontrivially is restricted by
something nontrivial a finite distance away from the boundary of the
moduli space, i.e., the singular loci resulting from blown-down curves
must be identified by the quotienting group in question.
In our examples, the
quotient singularities are naturally interpreted as fixed points of
the Weyl reflection of the root lattices of the $a, d, e$ Lie algebras,
thus their resolution will provide more possibilities of phase
transitions, perhaps leading to more realistic pairs.

It is also worth mentioning that most of the examples in this letter
can be generalized to the orientifold construction, using e.g, the
automorphisms on $T^2$ given by (\ref{ori1}), (\ref{ori2}), together
with reversion of worldsheet helicity. But as there are
no new insights in the construction, we will not report the result here.

\vskip 8mm
\noindent{\Large \bf Acknowledgements}
\vskip 2mm

Parts of the work have been done while the author was at the University
of Durham. I would like to thank Dr David Fairlie for support and kind
hospitality and the Royal Society for a K C Wong research fellowship.
This work is partially supported by the FWF under project no. P10641-PHY.

%\newpage

\eject


\newpage
\begin{thebibliography}{s2}

\bibitem{sen} A. Sen, "String-String Duality Conjecture in Six Dimensions
and Charged Solitonic Strings", Nucl. Phys. {\bf B450}(1995)103,
 hep-th/9504027.
\bibitem{hs} J.A. Harvey and A. Strominger, "The Heterotic String is a
Soliton", Nucl. Phys. {\bf B449}(1995)535, hep-th/9504047.
\bibitem{hult} C. Hull and P. Townsend, Nucl. Phys. {\bf B438}(1995)109.
\bibitem{kacvaf} S. Kachru and C. Vafa, "Exact Results for N=2
Compactifications of Heterotic Strings", hep-th/9505105, Nucl. Phys.
{bf B450}(1995)69.
\bibitem{2ndqu} S. Ferrara, J.A. Harvey, A. Strominger and C. Vafa,
"Second-Quantized Mirror Symmetry", Phys. Lett. {\bf 361B}(1995)59,
 hep-th/9505162.
\bibitem{witva} C. Vafa and E. Witten, "Dual String Pairs with $N=1$ and
$N=2$ Supersymmetry in Four Dimensions", hep-th/9507050.
\bibitem{senva} A. Sen and C. Vafa, "Dual Pairs of Type II String
Compactification", hep-th/9508064.
\bibitem{n1}J.A. Harvey, D.A. Lowe and A. Strominger, "N=1 String Duality,"
hep-th/9507168.
\bibitem{aspmor} P.S. Aspinwall and D. Morrison, "U-Duality and
Integral Structures", hep-th/9505025, Phys. Lett. {\bf 355B}(1995)141.
\bibitem{aspin2} P.S. Aspinwall, "Some Relationships Between Dualities in
String Theory", hep-th/9508154.
\bibitem{aspin1} Paul S. Aspinwall, "An N=2 Dual Pair and a Phase
Transition",  hep-th/9510142 .
\bibitem{chaulow} S. Chaudhuri and D.A. Lowe, "Type IIA-Heterotic Duals with
Maximal Supersymmetry", hep-th/9508144.
\bibitem{afiq} G. Aldazabal, A. Font, L.E. Ib\'a\~nez and F. Quevedo,
"Chains of N=2, D=4 Heterotic/Type II Duals", hep-th/9510093.
\bibitem{kac2}S.Kachru, A. Klemm, W. Lerche, P. Mayr and C. Vafa,
"Nonperturbative Results on the Point Particle Limit of the N=2
Heterotic String Vacua", hep-th /9508155
\bibitem{seiwit} N. Seiberg and E. Witten, Nucl. Phys. {\bf B426}
(1994)19; {\bf B431}(1994)484
\bibitem{klm} A. Klemm, W. Lerche and P. Mayr, "K3-Fibrations and
Heterotic-Type II String Duality", hep-th/9506112, Phys. Lett. {\bf
357B}(1995)313.
\bibitem{klt} V. Kaplunovsky, J. Louis and S. Theisen, "Aspects of
Duality in N=2 String Vacua," hep-th/9506110, Phys. Lett.
{\bf 357B}(1995)71.

\bibitem{liany} B.H. Lian and S.T Yau, "Mirror Maps, Modular Relations and
Hypergeometric Series I, II", hep-th/9507151, hep-th/9507153.
\bibitem{asplo} P. S. Aspinwall and J. Louis, "On the Ubiquity of K3
Fibration in String Duality", hep-th/9510234.
\bibitem{aspin3} P.S. Aspinwall, "Enhanced Gauge Symmetries and K3
Surfaces", Phys. Lett. {\bf 357}(1995)329, hep-th/9507012.
\bibitem{aspin4} P.S. Aspinwall, "Enhanced Gauge Symmetries and
Calabi-Yau Threefolds", hpe-th/9511171.

\bibitem{swsen} J. Schwarz and A. Sen, "The Type IIA Dual of the
Six-Dimensional CHL Compactification", hep-th/9507027.
\bibitem{niku} V.V. Nikulin, "Finite Automorphism Groups of K\"ahler
K3 Surfaces", Trans. Moscow Math. Soc. {\bf 38}(1979)71.
\bibitem{bart} W. Barth, "Lectures on K3 and Enriques Surfaces",
in {\it Algebraic Geometry, Sitges 1983}, Lect. Notes in Math. {\bf 1124},
Eds. E. Casas-Alvero, et al (Springer-Verlag, Berlin, Heidelberg 1985).
%\bibitem{wit} E. Witten, "String Theory Dynamics in Various Dimensions",
%Nucl. Phys. {\bf B}, hep-th/9503124
\bibitem{dstr} M. Bershadsky, C. Vafa and V. Sadov, "D-Strings on
D-Manifolds", hep-th/9510225.

\bibitem{wit2} E. Witten, "Some Comments on String Dynamics",
hep-th/9507121.

%I. Antoniadis, S. Ferrara, E. Gava, K.S. Narain and
%T.R. Taylor, Nucl. Phys. {\bf B447}(1995)35;
%B. de Wit, V. Kaplunovsky, J. Louis and D. L\"ust, Nucl. Phys.
%{\bf B451}(1995)53 and references therein.

\end{thebibliography}
\end{document}